\documentclass[3p,times]{elsarticle}

\usepackage{ecrc}

\volume{00}

\firstpage{1}

\journalname{Nuclear Physics A}

\runauth{Xin-Nian, Wang and Yan Zhu}

\jid{npa}

\jnltitlelogo{Nuclear Physics A}

\usepackage{amssymb,amsmath}

\biboptions{square,comma,numbers,sort&compress}

\usepackage[figuresright]{rotating}
\usepackage{endnotes}

\def\footnotetext{\endnotetext[\number\numexpr\value{endnote}+1]}
\let\footnotemark\endnotemark 

\def\enoteheading{\par\kern2\baselineskip%
                  \footnoterule%
                  \kern1\baselineskip}
\let\footnote=\endnote

\newcommand{\pt}{p_T}

\newcommand{\ba}{\begin{eqnarray}}
\newcommand{\ea}{\end{eqnarray}}

\newcommand{\fig}{Fig.~}

\newcommand{\gm}{\gamma}

\begin{document}

\begin{frontmatter}



\dochead{}

\title{Jet quenching and $\gamma$-jet correlation in\\
high-energy heavy-ion collisions\footnote{\vspace{-0.5cm}Talk presented by Y. Zhu.}}


\author[a1,a2]{Xin-Nian Wang}
\author[a3]{Yan Zhu}

\address[a1]{Key Laboratory of Quark and Lepton Physics (MOE) and Institute of Particle Physics,
Central China Normal University, Wuhan 430079, China}
\address[a2]{Nuclear Science Division Mailstop 70R0319, Lawrence Berkeley National Laboratory, Berkeley, California 94740, USA}
\address[a3]{Departamento de F\'{i}sica de Part\'{i}culas and
IGFAE, Universidade de Santiago de Compostela,
E-15706 Santiago de Compostela, Galicia, Spain}

\begin{abstract}
Medium modification of $\gm$-tagged jets in high-energy heavy-ion collisions is investigated within a linearized Boltzmann transport model which includes both elastic parton scattering and induced gluon emission. In Pb+Pb collisions at $\sqrt{s}=2.76$ TeV, a $\gm$-tagged jet is seen to lose 15\% of its energy at 0-10\% central collisions. Simulations also point to a sizable azimuthal angle broadening of $\gm$-tagged jets at the tail of a distribution which should be measurable when experimental errors are significantly reduced. An enhancement at large $z_\text{jet}=p_L/E_{\text{jet}}$ in jet fragmentation function at the Large Hadron Collider (LHC) can be attributed to the dominance of leading particles in the reconstructed jet. A $\gm-$tagged jet fragmentation function is shown to be more sensitive to jet quenching, therefore a better probe of the jet transport parameter. 
\end{abstract}


\end{frontmatter}


\section{Introduction}

Strong jet quenching \cite{Wang:1991xy} is caused by energy loss of energetic partons traversing through the strongly-interaction medium in high energy heavy-ion collisions. It leads to suppression and $p_T$ broadening of high $p_T$ hadrons and jets in the final states, as well as the modification of $\gamma-$jet and dijet correlations in A+A collisions with respect to p+p collisions. The energy loss is the result of multiple scattering and medium induced gluon radiation during the propagation of jets in the medium. In the work presented here, jet quenching and $\gm-$tagged jet modification in heavy-ion collisions at the LHC are studied within a linearized Boltzmann transport (LBT) model \cite{Li:2010ts,Wang:2013cia}. 

Since photons do not participate in strong interaction, $\gamma$-jets are excellent probes for the study of jet quenching without the geometrical trigger bias as in dijet correlations. 
Jet energy loss and jet structure can be calculated by reconstructing a jet with a jet-finding algorithm. In this study, the anti-$k_t$ algorithm within \textsc{fastjet} \cite{Cacciari:2011ma} is used.
In this presentation,  we will first give a brief introduction to the LBT model. By comparing with experimental data, we found a large energy loss for the $\gm-$tagged jets in Pb+Pb collisions at $\sqrt{s}=2.76$ GeV. We will also examine the azimuthal angle broadening and modification of jet fragmentation functions especially at large and small $z_\text{jet}$ .

In the LBT model, we assume that all partons propagate along classical trajectories between two adjacent collisions. The scattering center is determined by the probability of scattering, $P_{a}=1-\exp[-\sum_j (\Delta x_j\cdot u)\sum_{b}\sigma_{ab}\rho_{b}(x_j)]$ for a parton $a$, with $\sigma_{ab}$ the
parton scattering cross section and $\rho_{b}$ the local medium parton density,
and the sum over time steps $\Delta t_j=\Delta x_j$ (in natural units) starts from the last scattering point.  Each scattering is simulated according to the Boltzmann transport equation,
$
p_1\cdot\partial f_1(p_1)=-\int dp_2dp_3dp_4 (f_1f_2-f_3f_4)|M_{12\rightarrow34}|^2
(2\pi)^4\delta^4(p_1+p_2-p_3-p_4),
$
where $dp_i=d^3p_i/[2E_i(2\pi)^3]$, $|M_{12\rightarrow34}|^2=C g^4(\hat s^2+\hat u^2)/(-\hat t+\mu^2_{D})^2$ is the square of elastic scattering amplitude in a small angle approximation with Debye mass $\mu_{D}$,  $\hat s$, $\hat t$, and $\hat u$ are Mandelstam variables,
$C=1$ (9/4) is the color factor for quark-gluon (gluon-gluon) scattering,
$f_i=1/(e^{p\cdot u/T}\pm1)$ $(i=2,4)$ are parton phase-space distributions in a thermal medium with local temperature $T$ and fluid velocity $u=(1, \vec{v})/\sqrt{1-\vec{v}^2}$, and $f_i=(2\pi)^3\delta^3(\vec{p}-\vec{p_i})\delta^3(\vec{x}-\vec{x_i}-\vec{v_i}t)$ $(i=1,3)$ are the parton phase-space densities before and after scattering. 
Both shower $(p_3)$ and recoiled medium partons $(p_4)$ after each scattering are followed by further scatterings in the medium. The back reaction in the Boltzmann transport is implemented by transporting the initial thermal partons $(p_2)$, denoted as ``negative'' partons, according to the Boltzmann equation. 

\begin{figure}[t]
\centerline{\includegraphics[width=10cm]{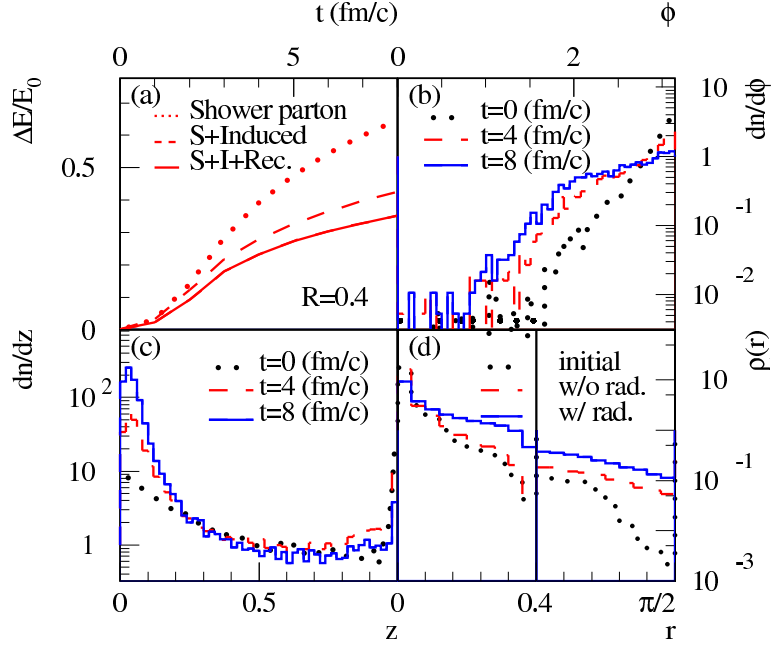}}
 \caption{ Energy loss as a function of time (a),
 azimuthal distribution relative to the $\gamma$ direction (b), jet fragmentation function (c) and jet profile (d)  for $\gamma$-tagged jets in a 
 uniform gluonic medium.
 \label{uniform}}
\end{figure}

To include induced radiation accompanying each elastic scattering, we using the higher-twist approach\cite{Wang:2001ifa},
$
\frac{dN_g^{a}}{dzdk_\perp^2dt}=\frac{6\alpha_{\rm s} P(z)}{\pi k_\perp^4} \frac{p\cdot u}{E} \hat{q}_{a} \sin^2\frac{t-t_i}{2\tau_f},
$
where $P(z)=[1+(1-z)^2]/z$ is the splitting function,  $\tau_f=2Ez(1-z)/k_\perp^2$ the formation time of the radiated gluon with energy fraction $z$ and transverse momentum $k_\perp$ emitted from a parent parton $a$ with energy $E$, and $\hat{q}_{a}=\sum_{b}\rho_{b}\int d\hat t q_\perp^2 d\sigma_{ab}/d\hat t $ the jet transport parameter. The Debye screening mass $\mu_D$ is used as an infrared cutoff for the gluon's energy. Multiple gluon emissions induced by a single  scattering are assumed to satisfy a Poisson distribution.
In the LBT model, all radiated gluons are assumed to be on-shell and their energies and momenta are successively 
determined from higher-twist approach. Interactions among radiated gluons, shower and recoiled partons are neglected. The strong coupling constant $\alpha_{\rm s}$ is fixed and will be determined via comparisons to experimental data.

\section{Results and Discussions}

Within the LBT model, we assume the medium excitation caused by jet-medium interaction to be small, $\delta f \ll f$ and neglect effects that are non-linear in $\delta f$. Initial jet shower partons are exported from \textsc{hijing} \cite{Wang:1991hta} by triggering a direct photon from a hard scattering with momentum transfer $q_t\ge30$ GeV in p+p collisions at $\sqrt{s}=2.76$ GeV.  $\gm-$tagged jets with $\pt^\gm>60$ GeV are selected for further propagation in LBT model. Using the anti-$k_t$ algorithm in \textsc{fastjet} \cite{Cacciari:2011ma}, the energy and momentum of a jet is reconstructed by summing up energy and momentum of all shower, radiated partons, and recoiled thermal partons, while subtracting those of negative partons inside the jet cone. It has been verified that there is only a very small difference between jet reconstructed in parton level and hadron level. Jet energy loss is shown in \fig \ref{uniform} (a) to increase as the propagation time in a uniform medium with temperature $T=300$ MeV and coupling constant $\alpha_{\rm s}=0.4$. Radiated gluons and recoiled partons are shown to take part of the jet energy. In \fig \ref{uniform} (b), jet quenching is seen to lead to a broadening of $\gm-$jet azimuthal angle distribution with $\phi=|\phi_\text{jet}-\phi_\gm|$. Jet fragmentation function $dn/dz(z=p_L/E_\text{jet})$ and jet profile $\rho(r)=\frac{1}{N_\text{jet}\Delta r}\sum_{\text{jets}}\pt(r-\Delta r,r+\Delta r)/\pt(0,R)$ are also shown in \fig\ref{uniform} (c) and (d). The fragmentation function clearly shows an enhancement at small $z$ while change at intermediate and large $z$ in a uniform medium. Jet transverse profile is broadened in and out of jet cone.

\begin{figure}
\centerline{\includegraphics[width=10.8cm]{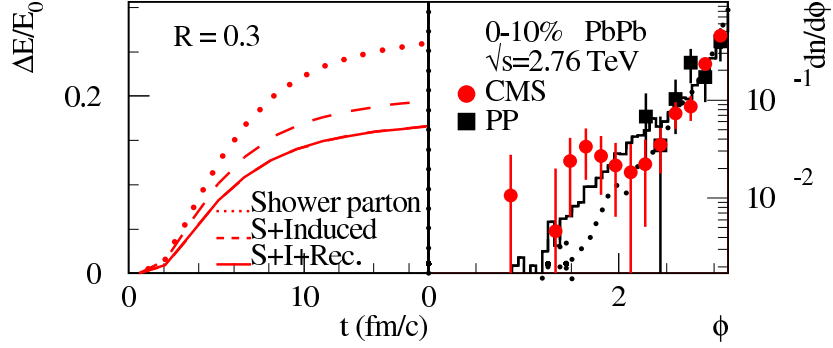}}
 \caption{Averaged energy loss as a function of time (left) and
 azimuthal distribution relative to the $\gamma$ (right) for $\gamma$-tagged jets
 in  central (0\%--10\%) Pb+Pb collisions at $\sqrt{s}=2.76$ TeV.  LBT results are obtained with $\alpha_s=0.2$ in a 3+1D hydro medium.
  \label{hydro}}
\end{figure}

\begin{figure}
\centerline{\includegraphics[width=10.3cm]{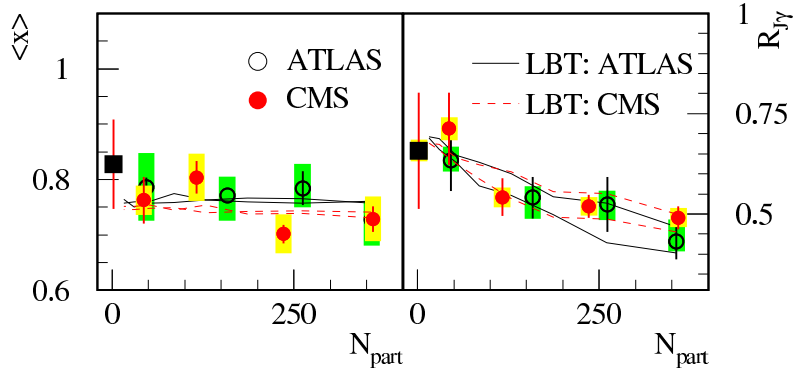}}
 \caption{Averaged $\gamma$-jet asymmetry $\langle x\rangle=\langle \pt^\text{jet}/\pt^\gm\rangle$  (left) and  jet survival rate $R_{J\gamma}$ (right) as functions of the number of participant nucleons in Pb+Pb collisions at $\sqrt{s}=2.76$ TeV from LBT as compared to
 experimental data  \cite{Chatrchyan:2012gt,ATLAS:2012cna}. Values of $\alpha_{\rm s}=0.15$--0.23 (dashed line) and 0.2--0.27 (solid line)
 are used for LBT calculations with CMS and ATLAS cuts, respectively.
 \label{avxr}}
\end{figure}

For comparison with experimental data for Pb+Pb collisions at the LHC, we further simulate propagation of $\gm-$tagged jets within LBT in a (3+1)D hydrodynamics \cite{Hirano:2005xf}. Initial $\gm-$jets from \textsc{hijing} are embedded in LBT according to the overlap function of two nuclei with a Wood-Saxon distribution. Different event selections are used according to experimental data from CMS and ATLAS. For CMS data \cite{Chatrchyan:2012gt},  $\pt^{\gamma}>60$ GeV, $|\eta^{\gamma}|<1.44$, $\pt^\text{jet}>30$ GeV, $|\eta^\text{jet}|<1.6$, and $\Delta\phi=|\phi^\text{jet}-\phi^\gm|>7\pi/8$, and for ATLAS data \cite{ATLAS:2012cna}, 
$60< \pt^{\gamma} < 90$ GeV, $|\eta^{\gamma}|<1.3$, $\pt^\text{jet}>25$ GeV, $|\eta^\text{jet}|<2.1$, and $\Delta\phi>7\pi/8$. In LBT simulations, we use a Debye screening mass $\mu^2_{D}=4\pi\alpha_{\rm s} T^2$.  \fig \ref{hydro} shows the average energy loss as a function of time (left), and $\gamma-$jet azimuthal angle distribution (right) with $\alpha_{\rm s}=0.2$. The energy loss similarly rises initially with time. It however saturates at $t\approx 10$ fm/$c$ because the jet transport parameter $\hat{q}_a=\hat{q}_a^0(\tau_0/\tau)^{1+\alpha}$ with $\alpha\ge 0$ decreases in an expanding medium with time and eventually terminates parton energy loss.  The $\gm-$jet azimuthal angle distribution (solid histograms) agrees well with the data (points) as shown in the right panel of \fig\ref{hydro}.  However, a small broadening relative to the p+p (dotted line) result in the large angle is clearly seen, though errors in the data are too big to confirm this. Therefore, more precise experimental measurements will be extremely useful to verify such an expected phenomenon.

In \fig \ref{avxr}, the averaged $\gamma-$jet asymmetry $\langle x\rangle=\langle \pt^\text{jet}/\pt^\gm\rangle$ and jet survival rate $R_{J\gamma}$ are compared with CMS \cite{Chatrchyan:2012gt} and ATLAS \cite{ATLAS:2012cna} data. The results are consistent with the data when fixed value of $\alpha_s=0.15$--0.23 for CMS and 0.2--0.27 for ATLAS is used, respectively. One can see that $\langle x\rangle$ is not quite sensitive to centrality while the jet survival rate is a better observable for measuring the centrality dependence of jet quenching.


\begin{figure}
\centerline{\includegraphics[width=10.8cm]{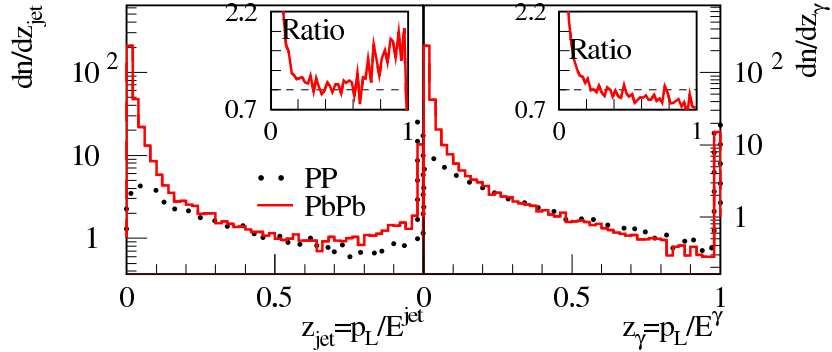}}
 \caption{The reconstructed (left) and $\gamma$-tagged jet fragmentation functions (right)
 within a jet cone $R=0.3$ of $\gamma$-tagged jets in central (0\%--10\%) Pb+Pb collisions at $\sqrt{s}=2.76$ TeV. 
 \label{frag}}
\end{figure}

Jet fragmentation function at LHC was shown to have an enhancement at both small and large $z_\text{jet}=p_L/E_\text{jet}$ \cite{CMS:2012wxa}. This phenomenon can be explained as the dominance of leading particle which takes a large energy fraction of a reconstructed jet. This is verified on the partonic level in \fig\ref{frag}. On the left panel, parton distribution function  in the fraction  of a reconstructed jet energy shows a similar property as the experimental observation. However, as shown in the right panel of \fig\ref{frag}, a $\gm$-tagged jet fragmentation function with $z_\gm=p_L/E_\gm$ does show an enhancement at small $z_\gm$ and a significant reduction at large $z_\gm$, which is more sensitive to jet medium interaction, and is a better probe to extract jet transport parameter.


We have reported jet quenching and $\gm-$jet correlations within a LBT model including multiple elastic scattering and induced radiation in a hot QCD medium. Jet quenching is observed as the energy loss of a reconstructed jet. The LBT results of $\gm-$jet azimuthal angle distribution, $\gm-$jet asymmetry and jet survival rate with fixed $\alpha_{\rm s}$ agree with CMS and ATLAS data quite well.  We observe, however, noticeable broadening in the $\gamma-$jet azimuthal angle distribution between Pb+Pb and p+p collisions in the large angle, which unfortunately cannot be discerned  by experimental data due to large errors. More precise measurements will be able to  verify this non-egligible broadening. The reconstructed jet fragmentation function from LBT exhibits the same behavior as the experimental observation, which shows enhancement at both small and large $z_\text{jet}$ due to the dominance of leading particle in a reconstructed jet. The $\gm$-tagged jet fragmentation function, however, is shown to be a more sensitive observable for jet quenching for a precise measurement of jet medium interact.

This work is supported by the NSFC under Grant No. 11221504,  China MOST under Grant No. 2014DFG02050, the Major State Basic Research Development Program in China (No. 2014CB845404), U.S. DOE under Contract No. DE-AC02-05CH11231 and within the framework of the JET Collaboration. Y.Z. was also supported by the the Alexander von Humboldt foundation, the DFG graduate school {\it{Quantum Fields and Strongly Interacting Matter}}, and European Research Council Grant No. HotLHC ERC-2001-StG-279579.

\vspace{-0.3cm}\theendnotes





\begin{thebibliography}{00}

\bibitem{Wang:1991xy} 
  X.~-N.~Wang and M.~Gyulassy,
  Phys.\ Rev.\ Lett.\  {\bf 68}, 1480 (1992).

\bibitem{Li:2010ts} 
  H.~Li, F.~Liu, G.-L.~Ma, X.-N.~Wang, and Y.~Zhu,
  Phys.\ Rev.\ Lett.\  {\bf 106}, 012301 (2011).
  
\bibitem{Wang:2013cia}
X.~-N.~Wang and Y.~Zhu,
Phys.\ Rev.\ Lett.\ {\bf 111}, 062301 (2013).

\bibitem{Cacciari:2011ma} 
  M.~Cacciari, G.~P.~Salam and G.~Soyez,
  Eur.\ Phys.\ J.\ C {\bf 72}, 1896 (2012).

\bibitem{Wang:2001ifa} 
  X.-N.~Wang and X.-F.~Guo,
  Nucl.\ Phys.\ {\bf A696}, 788 (2001).

\bibitem{Wang:1991hta} 
  X.-N.~Wang and M.~Gyulassy,
  Phys.\ Rev.\ D {\bf 44}, 3501 (1991).
  
  
  
  
\bibitem{Hirano:2005xf} 
  T.~Hirano, U.~W.~Heinz, D.~Kharzeev, R.~Lacey, and Y.~Nara,
  Phys.\ Lett.\ B {\bf 636}, 299 (2006).


\bibitem{Chatrchyan:2012gt} 
  CMS Collaboration,
  Phys.\ Lett.\ B {\bf 718}, 773 (2012).
  
\bibitem{ATLAS:2012cna} 
  ATLAS Collaboration, Report No.
  ATLAS-CONF-2012-121.
  
\bibitem{CMS:2012wxa}
  CMS Collaboration, Report No.
  CMS-PAS-HIN-12-013.

 \end{thebibliography}



\end{document}